\documentclass[aps,showpacs,showkeys,superscriptaddress,nofootinbib,preprint,11pt]{revtex4}
\usepackage{graphicx}% Include figure files
\usepackage{subfigure}

\usepackage{booktabs}
\usepackage{multirow}
\usepackage{subfigure}
\usepackage{graphicx}% Include figure files
\def\slashchar#1{\setbox0=\hbox{$#1$}
   \dimen0=\wd0 \setbox1=\hbox{/} \dimen1=\wd1
   \ifdim\dimen0>\dimen1 \rlap{\hbox to \dimen0{\hfil/\hfil}} #1
   \else  \rlap{\hbox to \dimen1{\hfil$#1$\hfil}} / \fi}

\newcommand{\dr}[1]{\multirow{2}{*}{#1}}
\newcommand{\psl}{\slashchar{p}}
\newcommand{\qsl}{\slashchar{q}}
\newcommand{\pksl}{\slashchar{p_k}}

\begin{document}

\title{Weak production of strange particles off the nucleon}

\author{M. Rafi Alam}
\affiliation{Department of Physics, Aligarh Muslim University, Aligarh-202 002, India}

\author{I. Ruiz Simo}
\affiliation{Dipartimento di Fisica, Universit\'a degli Studi di Trento
Via Sommarive 14, Povo (Trento)
I-38123, Italy}

\author{M. Sajjad Athar}
\affiliation{Department of Physics, Aligarh Muslim University, Aligarh-202 002, India}

\author{L. Alvarez-Ruso}
\affiliation{Departamento de F\'\i sica Te\'orica and Instituto de F\'isica Corpuscular, Centro Mixto
Universidad de Valencia-CSIC, E-46071 Valencia, Spain}

\author{M. J. Vicente Vacas}
\affiliation{Departamento de F\'\i sica Te\'orica and Instituto de F\'isica Corpuscular, Centro Mixto
Universidad de Valencia-CSIC, E-46071 Valencia, Spain}

\begin{abstract}
The strange particle production off the nucleon induced by 
neutrinos and antineutrinos is investigated at low and intermediate energies.
We develop a microscopic model based on the SU(3) chiral Lagrangian.
The studied mechanisms are the main source of single kaon production for 
(anti)neutrino energies up to 1.5~GeV. Using this model we have also studied the associated 
production of kaons and hyperons. 
The cross sections are large enough to be measured by experiments such as MINER$\nu$A, T2K and NO$\nu$A. 
\end{abstract}

\maketitle
%%%%%%%%%%%%%%%%%%%%%%%%%%%%%%%%%%%%%%%%%%%%
%% MAINMATTER
%%%%%%%%%%%%%%%%%%%%%%%%%%%%%%%%%%%%%%%%%%%%
\section{Introduction}
With the observation of a large $\theta_{13}$, neutrino experiments now pursue the discovery of CP violation in the leptonic sector, and the precise determination of neutrino mixing parameters. These goals require a better precision in the measurements, which can only be achieved with a very good theoretical and experimental understanding of neutrino interactions on both nucleons and nuclear targets. 
Not only the study of neutrino induced weak interactions is important to understand the analysis of the various
oscillation experiments, but would also help to unravel the structure of weak hadronic currents, in particular the strange quark content of the nucleon, provide better estimates of atmospheric backgrounds for nucleon decay searches and might even show hints for non-standard interactions. In the last several years,  $\nu-N$ quasi-elastic and one pion production processes have been studied in detail~\cite{Morfin:2012}. However, not much attention has been paid to other inelastic channels like single kaon production, associated strange particle production and single hyperon production~\cite{Rafi:2011}.

In neutrino induced reactions, the first inelastic process creating strange quarks is the single kaon ($K^+$ or $K^0$) production. This  charged current (CC) $|\Delta S|=1$ process is particularly appealing for several reasons. One of them is the important background that it could produce, due to atmospheric neutrino interactions, in the analysis of proton decay experiments. 
Another reason is its simplicity from a theoretical point of view. At low energies, it is possible to obtain model independent 
predictions using Chiral Perturbation Theory ($\chi$PT) and, due to the absence of $S=1$ baryonic resonances
 at these energies, the range of validity of the calculation could be reliably extended to energies higher than 
for other channels. Furthermore, the associated production of kaons and hyperons has a higher energy threshold (1.10 vs. 0.79 GeV) for $\nu_\mu$ induced processes. This implies that, even when the associated production is not Cabibbo suppressed, for a relatively wide energy region single kaon production could still be dominant. Similarly, in the case of antineutrino induced processes on nucleon targets, the threshold for associated antikaon production corresponds to the $K-\bar K$ channel, and is much higher than for the single kaon case. Therefore, a $|\Delta S|=1$ process, where an antikaon ($K^-$ or $\bar{K^0}$) is produced is the dominant source of antikaons for a wide range of antineutrino energies. Our study may be useful in the analysis of antineutrino experiments at  MINER$\nu$A, NO$\nu$A, T2K and others. For instance, MINER$\nu$A has plans to investigate 
several strange particle production reactions with both neutrino and antineutrino beams~\cite{Solomey:2005rs} with high statistics. 
Furthermore,  the T2K experiment~\cite{Kobayashi:2005}, LBNE experiment~\cite{LBNE} as well as beta beam experiments~\cite{Mezzetto:2010} will work at energies where the single (anti)kaon production may be important.

Here we present the results of our calculations for (anti)neutrino induced single (anti)kaon production~\cite{Rafi:2010,Alam:2012zz} reactions, and also for the associated production ($|\Delta S|=0$) channel. Our microscopic model is based on the SU(3) chiral Lagrangian. The basic parameters of the model are $f_\pi$, the pion decay constant, Cabibbo's angle, the proton and neutron magnetic moments and the  axial vector coupling constants for the baryons octet, $D$ and $F$, that are obtained from the analysis of the semileptonic decays of neutron and hyperons. For antineutrino induced single $K^-$ or $\bar{K^0}$ production, we have also considered a mechanism with the $\Sigma^*(1385)$ resonance in the intermediate state. In Sect.II, we describe the formalism in brief and in Sect.III, the results and discussions are presented.

\section{Formalism}
The basic reactions for $\nu(\bar\nu)$ CC single kaon production and associated particle production accompanied by a kaon from a nucleon ($p$ or $n$) target are,
\begin{equation}\label{S1:eq}
\left.
 \begin{array}{l} 
 \nu_{l} + N \rightarrow l^- + N^\prime + K \\
\bar\nu_{l}+ N \rightarrow l^+ + N^\prime + \bar K 
\end{array} \right\} \qquad \rm{CC} \quad (\Delta S =1 )
\end{equation}
 \begin{equation}\label{S3:eq}
\left.
 \begin{array}{l} 
 \nu_{l} + N \rightarrow l^- + Y + K \\
\bar\nu_{l}+ N \rightarrow l^+ + Y + K
\end{array} \right\} \quad \rm{CC}  \quad (\Delta S =0)
\end{equation}
where $l=e,\mu$, $ N \& N^\prime $ are nucleons and $Y$ denotes a hyperon. The expression for the differential cross section in lab frame for the above processes is given by
\begin{eqnarray}\label{d9_sigma}
d\sigma &=& \frac{(2\pi)^{4} }{4 M E} \prod_j \frac{d{\vec k_j}}{(2\pi)^{3}   2 E_{j} } 
\delta^{4} \left( \sum_i k_i - \sum_j k_j \right) \bar\Sigma\Sigma | \mathcal M |^2.
\end{eqnarray}
Here, $k_{i(j)}=(E_{i(j)},\vec{k}_{i(j)})$ are the initial(final) 4-momenta and $M$ is the nucleon mass. 
$ \bar\Sigma\Sigma | \mathcal M |^2  $ is the square of the transition amplitude averaged (summed) over the spins of the initial (final) state. At low energies, this amplitude can be written as
\begin{equation}
\label{eq:Gg}
 \mathcal M = \frac{G_F}{\sqrt{2}}\, j_\mu^{(L)} J^{\mu\,{(H)}}=\frac{g}{2\sqrt{2}}j_\mu^{(L)} \frac{1}{M_W^2}
\frac{g}{2\sqrt{2}}J^{\mu\,{(H)}},
\end{equation}
where $j_\mu^{(L)}$ and $  J^{\mu\,(H)}$ are the leptonic and hadronic currents respectively, 
$G_F$ is the Fermi coupling constant and 
$g$ is the gauge coupling.
The leptonic current can be readily obtained from the standard model Lagrangian coupling  of the $W$ bosons to the leptons 
\begin{equation}
{\cal L}=-\frac{g}{2\sqrt{2}}\left[{ W}^+_\mu\bar{\nu}_l
\gamma^\mu(1-\gamma_5)l+{ W}^-_\mu\bar{l}\gamma^\mu
(1-\gamma_5)\nu_l\right].
\end{equation}
The hadronic current is obtained using Chiral Perturbation Theory ($\chi$PT). 
The lowest-order SU(3) chiral Lagrangian describing the pseudo scalar mesons 
in the presence of an external weak current is discussed in Refs.~\cite{Rafi:2010,Alam:2012zz}, 
and is given by

\begin{figure}
\includegraphics[height=.23\textheight,width=0.8\textwidth]{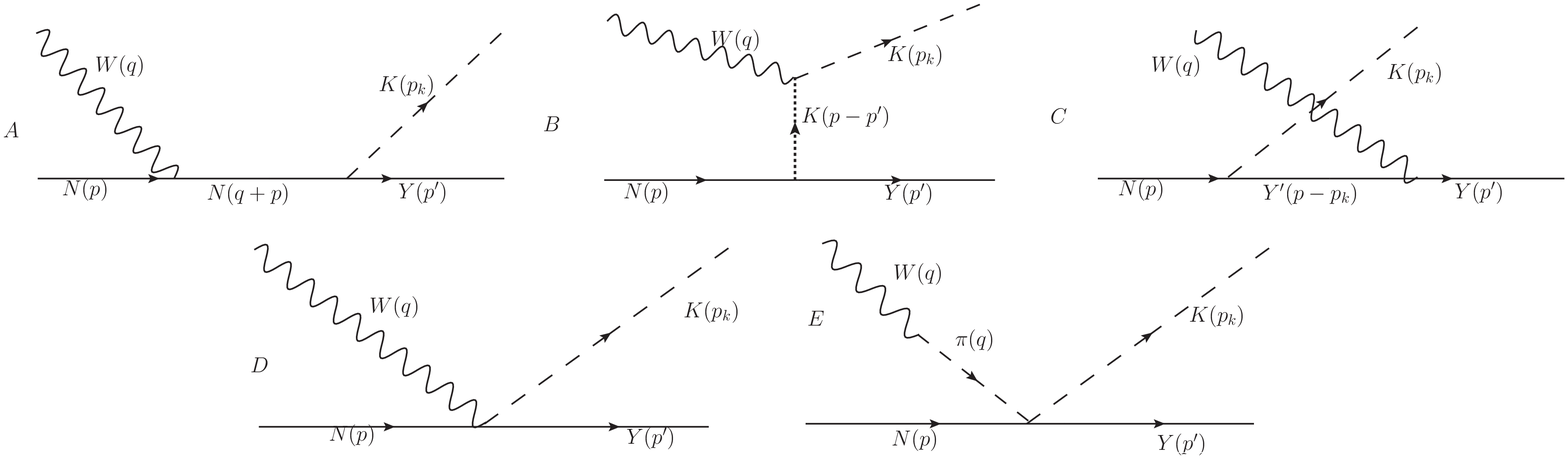}
\caption{Feynman diagrams for the $\nu / \bar \nu (k)  +   N (p) \rightarrow l (k^\prime) + K (p_k) + Y ( p^\prime )$ }
\label{fig:feynman}
\end{figure}

\begin{equation}
\label{eq:lagM}
{\cal L}_M^{(2)}=\frac{f_\pi^2}{4}\mbox{Tr}[D_\mu U (D^\mu U)^\dagger]
+\frac{f_\pi^2}{4}\mbox{Tr}(\chi U^\dagger + U\chi^\dagger),
\end{equation}
where the parameter $f_\pi=92.4$~MeV is the pion  decay constant, $U$ is the SU(3) representation of the meson fields
and its covariant derivative $D_\mu U = \partial_\mu U -i r_\mu U+iU l_\mu$. 
In the CC case, the right and left handed currents, $r_\mu=0$ and $l_\mu=-(g/\sqrt{2})( W^+_\mu T_+ + W^-_\mu T_-)$, where $W^{\pm}$ are the W fields while the matrices $T_\pm$ define the mixing of quark flavors and depend on the Cabibbo angle.
The lowest-order  chiral Lagrangian for the baryon octet in the presence of an external current can be written as~\cite{Rafi:2010}:
\begin{equation}
\label{eq:lagB}
{\cal L}_{MB}^{(1)}=\mbox{Tr}\left[ \bar B  \left( i \slashchar{D} 
-M_B \right) B \right]
-\frac{D}{2} \mbox{Tr}\left( \bar B \gamma^\mu \gamma_5 \{ u_\mu,B \} \right)
-\frac{F}{2}\mbox{Tr}\left( \bar B \gamma^\mu\gamma_5 \left[ u_\mu,B \right] \right),
\end{equation}
where $M_B$ denotes the mass of the baryon octet and $B$ is the Baryon SU(3) matrix; $D_\mu B = \partial_\mu B + [\Gamma_\mu,B]$ where $\Gamma_\mu$ depends on the meson fields and the external currents $r_\mu$ and $l_\mu$ introduced above.
The parameters $D=0.804$ and $F=0.463$ which are determined from the baryon semileptonic decays. 
For the single kaon/antikaon production, the Feynman diagram and the corresponding hadronic currents are given in
Refs.~\cite{Rafi:2010,Alam:2012zz}. Here we are discussing explicitly the $\Delta$S=0 channel. The Feynman diagrams contributing in this case are shown in Fig.~\ref{fig:feynman}.
The first row consists of s-channel, t-channel and u-channel diagrams, whereas the next row has the contact and $\pi$-pole terms. Using the chiral Lagrangian for the different interaction vertices, the different contributions to the hadronic current coupling to the $W$ bosons  for the above set of Feynman diagrams are
\begin{table}
%\begin{center}
\begin{tabular}{|l|ccccccc|} \hline
Process 	                                & $A_{CT}$	            &$B_{CT}$		      &     $A_{SY}$  	             & $A_{U\Sigma}$                 & $A_{U\Lambda}$      & $A_{TY}$                     &  $A_{\pi }$ \\  \hline \hline 
$\bar\nu_l  p \rightarrow l^+  \Lambda  K^0 $ &\dr{$-\sqrt{\frac32}$}       &\dr{$\frac{-1}{3}(D+3F)$}&\dr{$\frac{-1}{\sqrt6}(D+3F)$}& \dr{ $-\sqrt{\frac23}(D-F)$ } &  \dr{  0}           &\dr{$\frac{-1}{\sqrt6}(D+3F)$}& \dr{$\sqrt{\frac32}$} \\
$\nu_l n \rightarrow l^- \Lambda  K^+  $ & & & & & & & \\ \hline
$ \bar\nu_l  p \rightarrow l^+  \Sigma^0  K^0 $ &\dr{$\mp \frac{1}{\sqrt2}$}&\dr{$D-F$}	              &\dr{$\mp\frac{1}{\sqrt2}(D-F)$}& \dr{$\mp \sqrt2 (D-F)$ }     &   \dr{0}            &\dr{$\pm\frac{1}{\sqrt2}(D-F)$}& \dr{$\pm\frac{1}{\sqrt2}$ }  \\ 
$\nu_l n \rightarrow l^-\Sigma^0 K^+ $  & & & & & & & \\  \hline %\midrule\specialrule{2.5pt}{1pt}{1pt}
$ \bar\nu_l  p \rightarrow l^+  \Sigma^- K^+ $  & \dr{0}		    &\dr{0}  		      &   \dr{ $D-F$	}	     & \dr{$D-F$}                    & \dr{$\frac13(D+3F)$}& \dr{0}         	          & \dr{ 0}	   \\
$\nu_l n \rightarrow l^-\Sigma^+ K^0 $  & & & & & & & \\  \hline
$ \bar\nu_l  n \rightarrow l^+  \Sigma^-  K^0 $ & \dr{$-1$}		    & \dr{$D-F$}	      &\dr{0}  		             & \dr{$F-D$}	             &\dr{$\frac13(D+3F)$} &   \dr{$D-F$ }     	          & \dr{ $1$ } \\
$\nu_l p \rightarrow l^-\Sigma^+ K^+ $  & & & & & & & \\ 
\hline
\end{tabular}
\caption{Constant factors  appearing in the hadronic current. The upper(lower) sign corresponds to the 
processes with $\bar \nu(\nu)$.}\label{tb:currents}
%\end{center}
\end{table}
\begin{eqnarray*}
j^\mu \arrowvert_{CT} &=&i A_{CT} V_{ud} \frac{ \sqrt{2}}{2 f_\pi} \bar u_Y(p^\prime) \; (\gamma^\mu + B_{CT} \; \gamma^\mu \gamma^5 ) \; u_N(p) \\
j^\mu \arrowvert_{SY} &=&i A_{SY} V_{ud} \frac{ \sqrt{2}}{2 f_\pi} \bar u_Y (p^\prime) \pksl \; \gamma_5  \frac{ \psl +
  \qsl + M}{( p +  q)^2 -M^2} \left(\gamma^\mu +i  \frac{(\mu_p - \mu_n)}{2 M} \sigma^{\mu \nu} q_\nu \right. \\
  && \left. - (D+F) \left\{ \gamma^\mu 
  - \frac{q^\mu}{ q^2-{m_\pi}^2 } \qsl \right\} \gamma^5 \right) u_N(p) \\
j^\mu \arrowvert_{UY} &=&i A_{UY'}V_{ud} \frac{ \sqrt{2}}{2 f_\pi} \bar u_Y(p^\prime) \, H^\mu_{Y^\prime} \, \frac{ \psl - \pksl + M_{Y'}}
{( p - p_k)^2 -M_{Y'}^2} \pksl \; \gamma_5  u_N(p) \\
j^\mu \arrowvert_{TY}&=& i A_{TY} V_{ud} \frac{\sqrt{2}}{2 f_\pi} (M + M_Y) \bar u_Y(p^\prime) \gamma^5 \; u_N(p) \frac{q^\mu - 2 p_k^\mu}{(q - p_k)^2-m_k^2}    \\
j^\mu \arrowvert_{\pi P} &=& i A_{\pi }V_{ud} \frac{\sqrt{2}}{4 f_\pi} \bar u_Y(p^\prime)(\qsl + \pksl \;) u_N (p) \frac{q^\mu}{q^2 - m_\pi^2} \\
\rm{with}&& \\
H^\mu_\Lambda & = & i \frac{3 \mu_n }{4M} \sigma^{\mu \nu} q_\nu + D \left( \gamma^\mu - \frac{ \qsl q^\mu}{q^2 - m_\pi^2} \right) \gamma^5 \\
H^\mu_\Sigma & = & \gamma^\mu +  i \frac{2 \mu_p + \mu_n }{4M} \sigma^{\mu \nu} q_\nu - F \left( \gamma^\mu - \frac{  \qsl   q^\mu}{q^2 - m_\pi^2}   \right) \gamma^5.
\end{eqnarray*}
Here $Y$ stands for the hyperon in the final state and $Y'$ for the one in the intermediate one in the u-channel diagram. As the dependence of the different terms on the momentum transferred to the nucleon is poorly known, if at all, a global dipole form factor $F(q^2) =(1 -q^2/M_A^2)^{-2}$ with a natural value of $M_A =1.05$~GeV has been taken for the numerical 
calculations. It's effect is small at relatively low neutrino energies.

\section{Results and Discussion}

\begin{figure}
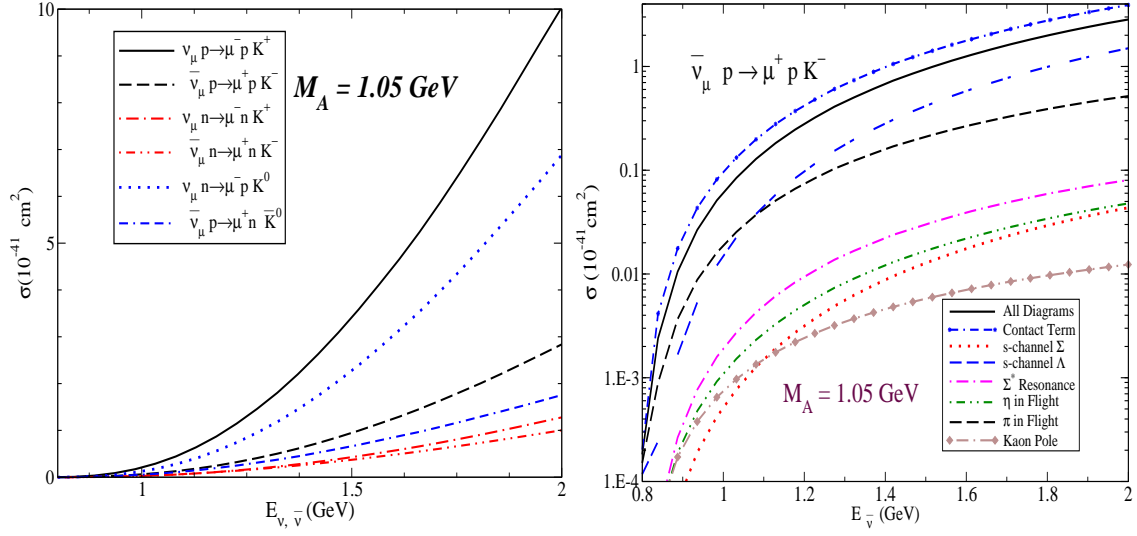

\includegraphics[height=.3\textheight,width=0.45\textwidth]{Total.eps}
\includegraphics[height=.3\textheight,width=0.45\textwidth]{PP_muon_log.eps}
\caption{Cross section for the $ | \Delta S | = 1 $  single kaon production.}
\label{fig:xsecKKbar}
\end{figure}

\begin{figure}[b]
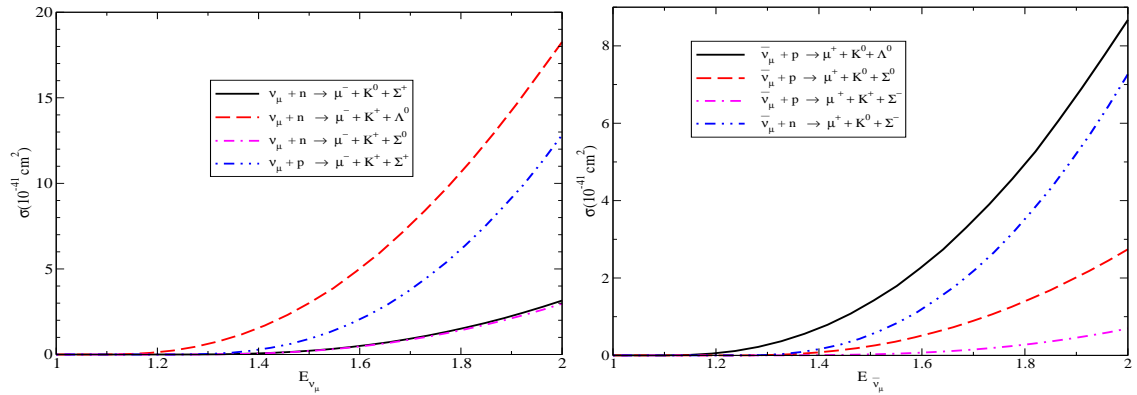

\includegraphics[height=.22\textheight,width=0.45\textwidth]{nu.eps}
\includegraphics[height=.22\textheight,width=0.45\textwidth]{nubar.eps}
\caption{Cross section for the $ | \Delta S | = 0$  associated kaon production.}
\label{fig:xsecap}
\end{figure}

The total scattering cross section $\sigma$ has been obtained from Eq.~(\ref{d9_sigma})
after integrating over the kinematical variables.
The cross sections for the $\nu_\mu N \rightarrow \mu^- N^\prime  K$ and  $\bar \nu_\mu N \rightarrow \mu^+ N^\prime \bar K$ 
processes are given in the left panel of Fig.~\ref{fig:xsecKKbar} whereas in the right panel we show the explicit 
dependence of the different terms of the amplitude for $\bar \nu_\mu p \rightarrow \mu^+ p^\prime K^- $  
\footnote{Further details can be found in our papers on kaon production \cite{Rafi:2010,Alam:2012zz}}.
It is clear from the figure that the contact term comes out to be the dominant one.
For the said process the other important terms are u-channel with a $\Lambda$ in the intermediate state and the $\pi$ in flight. 
We have investigated all the channels for single kaon production
and they have almost the same trend \cite{Rafi:2010,Alam:2012zz}.
Here it is important to note that the resonant term is not so significant even at high energies, in variance with 
weak pion production, where the cross section comes predominantly from the $\Delta(1232)$ resonance at the energies 
considered here.

Unlike the single kaon production channels, $\Delta S = 0$ processes are not Cabibbo suppressed. Therefore, they are assumed to be dominant even at low energies.  Moreover, there may be important resonant contributions to the $\Delta S = 0$ processes, which we have not considered in the present study. However, we must point out that the inclusion of these resonances, which make the dynamics of these channels more involved, can play a relevant role as it is known from associated strangeness photoproduction studies. Recently, a model that describes reasonably $\pi, \gamma \rightarrow K \Lambda, K \Sigma$ has been used to predict the corresponding reactions induced by neutrinos in the forward direction by applying PCAC~\cite{Kamano:2012id}.

Within our model the contact term turns out to be the most dominant contribution followed by the s- and u- channel diagrams. Using Eq.~(\ref{d9_sigma}), we have obtained the cross section as a function of the (anti)neutrino energy. The results are shown in Fig.~\ref{fig:xsecap} for neutrino (left panel) and antineutrino (right panel) induced reactions.
We find that the cross sections for the reaction channels with a $\Lambda$ in the final state are the largest. This can be understood from the relative strength of the coupling $g_{NK\Lambda} = \sqrt{3} (D + 3 F)/(6 f_\pi)$ vs $g_{NK\Sigma}=-3 (D-F)/(6 f_\pi)$. Furthermore, $\Lambda$ production is favored by the available phase space due to its small mass relative to $\Sigma$ baryons. For $\nu_\mu n \rightarrow \mu^- \Sigma^+ K^0$ and $\bar \nu_\mu p \rightarrow \mu^+ \Sigma^- K^+ $ there is no contribution from the contact term and hence the cross sections are relatively lower.

\begin{figure}[t]
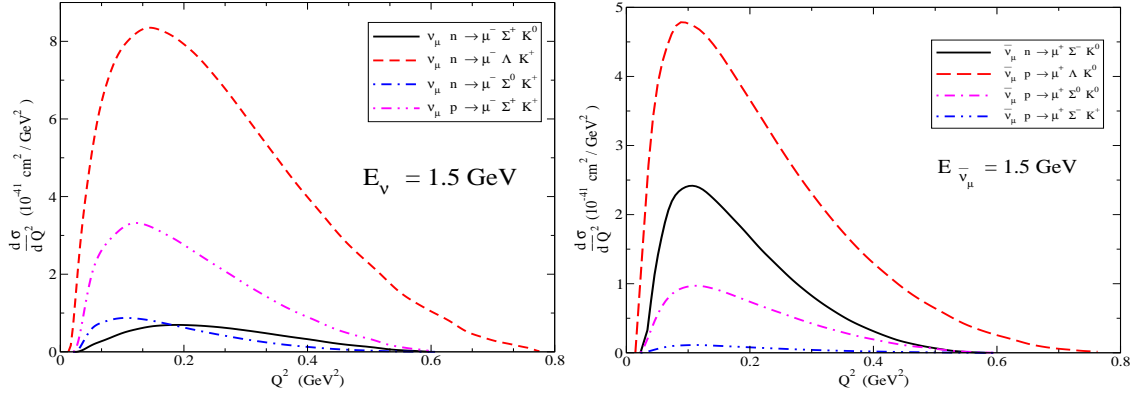

\includegraphics[height=.22\textheight,width=0.45\textwidth]{Q2_1_5_nu.eps}
\includegraphics[height=.22\textheight,width=0.45\textwidth]{Q2_1_5_nubar.eps}
\caption{$Q^2$ distribution for $|\Delta S| =0$ channels.}
\label{fig:q2ap}
\end{figure}

We have also studied the nature of the $Q^2$ distribution near threshold at 1.5 GeV for (anti)neutrino induced $\Delta S = 0$ processes and the results are shown in Fig.~\ref{fig:q2ap}. We also find that the processes with a $\Lambda$ in the 
final state have a sharper peak than the other channels, while $\nu_\mu n \rightarrow \mu^- \Sigma^+ K^0$ has a broader peak. 
 It is also important to note that the $Q^2$ distributions for antineutrino reactions are more forward peaked than those of the corresponding neutrino channels. 

To summarize, in this work we have studied single kaon production and the associated strange particle production processes on nucleons induced by (anti)neutrinos. The present study, based on the SU(3) chiral Lagrangians, should be quite reliable at low and intermediate energies as the parameters of the model are well known. For the single kaon production, we find that the cross sections are large enough to be measured at the Minerva, T2K and other experiments. Furthermore, the study of these cross sections may be helpful in estimating background contribution in nucleon decay searches.
 The present study of associated particle production is a first step towards a realistic description of these reactions that would be helpful to update the present models used in neutrino Monte Carlo generators.  

\section{ACKNOWLEDGMENTS}
MRA and MSA are thankful to the Aligarh Muslim University for the financial support to attend NuInt12. This work is partially supported by the Spanish Ministerio de Econom\'ia y Competitividad and European FEDER funds
under Contracts FIS2011-28853-C02-01 and  FIS2011-28853-C02-02, Generalitat Valenciana under contract PROMETEO/2009/0090 and the EU Hadron-Physics2 project, Grant No. 227431.

\end{document}